\documentclass[twocolumn,showpacs]{revtex4}

\usepackage{epsfig}
\usepackage{graphicx}
\usepackage{dcolumn}
\usepackage{bm}
\usepackage{hyperref}
\usepackage{verbatim}
\usepackage{multirow}
\newcounter{qcounter}
\usepackage{amsmath} 
\usepackage{textcomp}
\usepackage{mathrsfs}

\newcommand{\od}[2]{\left( #1 \right)_{ #2 }}

\begin{document}
\addtocounter{qcounter}{1}

\title{Alternatives to standard puncture initial data for binary black hole
evolution}

\author{George Reifenberger}
\affiliation{Department of Physics, Florida Atlantic University,
             Boca Raton, FL  33431}

\author{Wolfgang Tichy}
\affiliation{Department of Physics, Florida Atlantic University,
             Boca Raton, FL  33431}

\pacs{
04.25.dg,	
04.25.Nx,	
04.30.Db,	
04.30.Tv,	
}

\begin{abstract}
Standard puncture initial data have been widely used for numerical binary black
hole evolutions despite their shortcomings, most notably the inherent lack of
gravitational radiation at the initial time that is later followed by a burst of
spurious radiation. We study the evolution of three alternative initial data
schemes. Two of the three alternatives are based on post-Newtonian expansions
that contain realistic gravitational waves. The first scheme is based on a
second order post-Newtonian expansion in Arnowitt, Deser, and Misner
transverse-traceless (ADMTT) gauge that has been resummed to approach standard
puncture data at the black holes. The second scheme is based on asymptotic
matching of the 4-metrics of two tidally perturbed Schwarzschild solutions to a
first-order post-Newtonian expansion in ADMTT gauge away from the black holes.
The final alternative is obtained through asymptotic matching of the 4-metrics
of two tidally perturbed Schwarzschild solutions to a second-order
post-Newtonian expansion in harmonic gauge away from the black holes. When
evolved, the second scheme fails to produce quasicircular orbits (and instead
leads to a nearly head-on collision). This failure can be traced back to
inaccuracies in the extrinsic curvature due to low-order matching. More
encouraging is that the latter two alternatives lead to quasicircular orbits and
show gravitational radiation from the onset of the evolution, as well as a
reduction of spurious radiation. Current deficiencies compared to standard
punctures data include more eccentric trajectories during the inspiral and
larger constraint violations, since the alternative data sets are only
approximate solutions of Einstein's equations. The eccentricity problem can be
ameliorated by adjusting the initial momentum parameters. 
\end{abstract}

\maketitle
\section{Introduction}
\label{Intro}
One of the best candidates for observation of gravitational radiation is the
coalescence of two black holes. The successful detection of these gravitational
waves by any ground or space-based interferometer (e.g. 
LIGO~\cite{LIGO:2007kva,LIGO_web},
VIRGO~\cite{VIRGO_FAcernese_etal2008,VIRGO_web}, GEO600~\cite{GEO_web},
eLISA/NGO~\cite{eLISA/NGO_web}) will be confirmed by matched filtering of the
observed signal against an extensive compilation of waveforms produced
numerically from the binary black hole parameter space. Therefore, it
is important that this collection of constructed waveforms represent the most
physically viable scenarios of binary black hole evolution.

Full numerical evolution of the Einstein equations made possible since
2005~\cite{Pretorius:2005gq,Campanelli:2005dd,Baker:2005vv} is the method
of choice for producing waveforms than span part of the inspiral, and the
subsequent merger and ringdown phases. The waveform templates are dependent on
the physical accuracy of the data being used. Currently, the most widely used
approach for initial data construction is known as the ``puncture'' method.
Developed by~\cite{Brandt97b}, this method is able to set up data for binary
black holes with generic momenta and spins.
These initial data are used almost exclusively by all groups that evolve
black holes using the BSSNOK formulation~\cite{Baumgarte:1998te} together
with the moving punctures approach~\cite{Campanelli:2005dd,Baker:2005vv}.
Standard puncture initial data have the disadvantage that they contain
no realistic gravitational waves, because a conformally flat metric is used
in this method. When evolved, these data lead to a gravitational wave signal
that is zero for some time followed by a spurious burst. Only after this
burst do we get an astrophysically realistic chirp signal, thereby making the
prescription inherently unphysical. Other methods have begun to emerge whereby
post-Newtonian (PN) approximations are used, either as the sole
contributor~\cite{Tichy02,Kelly:2007uc,Kelly:2009js} or in conjunction with
analytical solutions near the black
hole~\cite{Yunes:2005nn,Yunes:2006iw,JohnsonMcDaniel:2009dq}, to develop
initial data. These methods allow for increased astrophysical accuracy by
including, to leading PN order, gravitational waves on the spacetime from the
onset of the simulation.
In this paper, we investigate three methods of initial data construction and
use them for full numerical evolution of binary black holes. These methods
can be considered alternatives to the standard puncture technique for
creating templates. We find that evolutions of two of these methods lead to
improved gravitational waves. There is a chirp signal right from the start,
and the (still present) burst of spurious radiation has a smaller amplitude than
for evolutions starting from standard puncture data.
The burst occurs when the built-in post-Newtonian waveform transitions to a
fully numerically generated signal.

Throughout this paper, we will use units where $G=c=1$. Latin indices such as
$i$ run from 1 to 3 and denote spatial indices, while Greek indices such as
$\mu$ run from 0 to 3 and denote spacetime indices. The paper is organized as
follows: In Section~\ref{3+1 split}, we briefly recall the 3+1 formulation for
numerical relativity. In Section~\ref{Standard Punctures}, we summarize the
standard puncture technique for creating initial data and discuss weaknesses in
this approach. In Section~\ref{ADMTT-PN}, we describe our first alternative
scheme; a post-Newtonian expansion of the 3-metric and extrinsic curvature in
Arnowitt, Deser, and Misner transverse-traceless (ADMTT) gauge~\cite{Schaefer85}
resummed to approach standard puncture data at the black holes. In
Section~\ref{Asymptotic Matching}, we describe the technique of asymptotic
matching the 4-metrics of tidally perturbed Schwarzschild black hole solutions
to post-Newtonian expansions away from the black holes. Also within this
section, we describe two more initial data schemes: asymptotic matching with a
low-order post-Newtonian expansion of the 4-metric in ADMTT gauge away from the
holes, and asymptotic matching of the 4-metrics with a second-order
post-Newtonian expansion in harmonic gauge away from the black holes. In
Section~\ref{Numerical_Explanantion_&_Results}, we describe additional
adjustments required for numerical evolution. In particular, we describe an
algorithm we use to fill black hole interiors for initial data that contain
singularities. We then show comparisons of evolutions starting from different
initial data schemes. Section~\ref{Discussion} summarizes our results and
includes discussion of future improvements to certain schemes.


\section{3+1 Formulation}
\label{3+1 split}
For numerical evolutions of general relativity, it is useful to split the
four-dimensional spacetime described in terms of the metric $g_{\mu\nu}$
into three-dimensional spatial hypersurfaces and time. The
Arnowitt-Deser-Misner~\cite{Arnowitt62} decomposition of the Einstein
equations defines the line element as
\begin{eqnarray}
 ds^{2} = -{\alpha}^{2}{dt}^{2} + g_{ij}({dx}^{i} +
{\beta}^{i})({dx}^{j}
+ {\beta}^{j}),
\end{eqnarray}
where $\alpha$ is the lapse function, ${\beta}^{i}$ is the shift vector, and
$g_{ij}$ is the 3-metric. The extrinsic curvature is defined as
\begin{eqnarray}
 K_{ij} = -{\frac{1}{2{\alpha}}}({\partial_t}g_{ij} -
{\mathcal{L}_{\beta}}g_{ij}).
\end{eqnarray}

Under the Arnowitt-Deser-Misner decomposition, the Einstein equations yield
evolution
equations of the form
\begin{eqnarray}
{\partial_t}g_{ij} 
&=& 
   -2{\alpha}K_{ij} + {\mathcal{L}_{\beta}}g_{ij}\\
{\partial_t}K_{ij} 
&=&
   {\alpha}(R_{ij} 
		  - 2K_{ij}K^{ij}) 
				  - D_iD_j\alpha
					        + {\mathcal{L}_{\beta}}K_{ij}
\nonumber \\
& & - 8{\pi}S_{ij} 
		  + 4{\pi}g_{ij}(S - {\rho})
\end{eqnarray}
and the Hamiltonian and momentum constraint equations
\begin{eqnarray}
R 
  - K_{ij}K^{ij}
	        + K^{2} 
&=&
   16{\pi}{\rho}\\
D_j(K^{ij}
	  - g^{ij}K)
&=&
   8{\pi}j^{i}.
\end{eqnarray}
Here, $R_{ij}$ and $R$ are the Ricci tensor and scalar of the 3-metric
$g_{ij}$, and $D_i$ is the derivative operator compatible with $g_{ij}$.
In binary black hole simulations, these equations simplify due to the
vanishing of the source terms $\rho$, $j^{i}$, $S_{ij}$, and
$S=g^{ij}{S_{ij}}$.


\section{Standard Puncture Initial Data}
\label{Standard Punctures}
The most common approach to producing initial data used currently is the
standard puncture prescription of Brandt \& Br\"{u}gmann~\cite{Brandt97b}
using Bowen-York extrinsic curvature. This method uses a conformal
transverse-traceless decomposition of the extrinsic
curvature~\cite{Baumgarte:2002jm,Cook00a} to simplify the constraint
equations. A conformally flat three-metric, $g_{ij}$, is defined as
\begin{eqnarray} \label{eq:g_BY}
g_{ij} = \psi^4 {\delta}_{ij},
\end{eqnarray}
where $\psi$ is the conformal factor. The extrinsic curvature  is given by
\begin{equation}
K^{ij} = \psi^{-10} A^{ij}_{BY},
\end{equation}
where 
\begin {eqnarray} \label{eq:A_BY}
A^{ij}_{BY}&=&\sum_{A=1}^2 \frac{3}{2r^{2}_{A}}
[p^{i}_{A}n^{j}_{A}+p^{i}_{A}n^{i}_{A}
-(g^{ij}-n^{i}_{A}n^{j}_{A})p^{k}_An_{Ak}] \nonumber \\
& & +\frac{3}{r^{3}_A} \left(\epsilon^{ikl}s_{Ak}n_{Al}n^{j}_A +
\epsilon^{jkl}s_{Ak}n_{Al}n^{i}_A\right),
\end {eqnarray}
is of Bowen-York form and where $n^{a}_{A}$ is the radial normal vector,
$p^{a}_{A}$ is the linear momentum, and $s^{a}_{A}$ is the spin of black hole
$A$. This $K_{ij}$ is chosen so that it already satisfies the momentum
constraint for arbitrary spin and momentum. 

The exact solution for a black hole pair with zero momenta and spins is the
Brill-Lindquist solution~\cite{Brill63}
\begin{eqnarray}
\label{eq:psiBL}
\psi_{BL} = 1 + \sum_{A=1}^2 \frac{m_A}{2r_A},
\end{eqnarray}
where $m_A$ is the ``bare'' puncture mass of black hole $A$ and
$r_A = |\vec{x} - \vec{x}_A|$ is the distance from the black hole
at position $\vec{x}_A$. As one can see, this conformal factor diverges at
the puncture location $\vec{x}_A$. Such divergences also occur for non-zero
masses and spins. To avoid numerical problems with these divergences when
solving the Hamiltonian constraint equation, one usually splits the conformal
factor~\cite{Brandt97b}
\begin{eqnarray}
\label{eq:psiPunc}
\psi = \psi_{BL} + u
\end{eqnarray}
into a regular piece $u$ and the divergent piece $\psi_{BL}$. By doing this,
the Hamiltonian constraint is reduced to a single elliptic equation 
\begin{equation}
\Delta^{flat}u =
-\frac{1}{8}{\psi_{BL}}^{-7}
                 A^{ij}_{BY} A^{kl}_{BY} \delta_{ik} \delta_{jl}
\left(1+\frac{1}{\psi_{BL}}u\right)^{-7}
\end{equation}
without any divergences, that can  be easily evaluated by elliptic solvers
such as in~\cite{Ansorg:2004ds}. Notice that $A^{ij}_{BY}=0$ for zero
momentum and spin, so that $u=0$. Furthermore, if one of the masses (e.g.
$m_2$) is zero we obtain the Schwarzschild solution in isotropic coordinates.

The fact that data are built using the assumption of conformal
flatness is an inherent problem. It is unrealistic to demand conformal
flatness in the context of astrophysical constructions, because it implies
that there are no gravitational waves on the initial time slice.

\begin{figure}[htp]
\includegraphics[scale=0.33,clip=true,angle=-90]{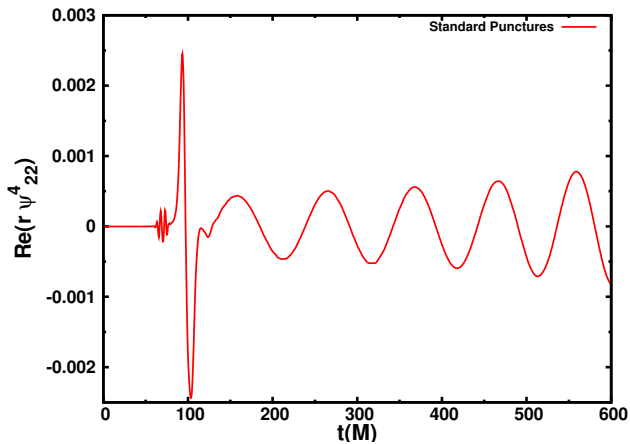}
\caption{\label{cyPunc_Re_rpsi4}
The real part of the dominant $\ell=2$, $m=2$ waveform mode of $\psi_4$ over
time extracted at a distance of $r=90M$. The lack of a gravitational wave
signature for the initial phase of the evolution is a direct consequence of
prescribing conformal flatness. The burst behavior (visible as the waveform
transitions from zero to a regular chirp signal) is indicative of a so-called
spurious radiation signal.
}
\end{figure}

Often, gravitational waveforms are described in terms of the Newman-Penrose
scalar, $\Psi_4$, extracted at some distance from the sources. Usually this
$\Psi_4$ is decomposed into individual $\ell$ and $m$ modes by projecting
onto spherical harmonics $Y{^{-2}_{{\ell}m}}$ of spin weight
-2~\cite{Brugmann:2008zz}. Figure~\ref{cyPunc_Re_rpsi4} shows the real part
of the dominant $\ell=2$, $m=2$ waveform mode of $\psi_4$ over time extracted
at a distance of $90M$. We observe a lack of any gravitational radiation
until $\sim 80M$; this delay in the waveform signal at arbitrary extraction
radii is common of standard punctures. Also, we see the beginning of an
outward traveling wave profile only after about $\sim 150M$ and differing
significantly from the initial burst; the waves up until this outward profile
region have been commonly referred to as spurious radiation.

If such a simulation is used to make gravitational wave templates, the early
part of the wave signal has to be discarded up to the time when the spurious
radiation has become sufficiently small. How much exactly one has to discard
depends on the desired accuracy, but in our experience it typically amounts
to between 10\% and 20\% of the total simulation time. This kind of waste
will only increase with future code upgrades that will allow us to extract
the waves at larger distances, since then we have to wait longer for the
spurious burst to even reach the extraction distance. These problems are the
reason why we seek alternative constructions of binary black hole initial
data. In~\cite{Lovelace:2008hd} it was shown that spurious radiation can be
reduced by using a 3-metric that is not conformally flat.

\section{Initial Data Based on Post-Newtonian Expansion in ADMTT Gauge}
\label{ADMTT-PN}
One alternative to standard puncture data has been introduced by Tichy
\textit{et al.} in~\cite{Tichy02} with further developments
in~\cite{Kelly:2007uc,Kelly:2009js}. This method uses post-Newtonian
expansions in ADMTT gauge~\cite{Schaefer85}
for the 3-metric and extrinsic curvature on the
initial time slice. While PN theory is strictly valid only away from regions
of strong gravity such as black holes, it is possible to rewrite the 3-metric
and extrinsic curvature in ADMTT gauge in such a way that they approach the
puncture form as in Eq.~\eqref{eq:g_BY} and Eq.~\eqref{eq:A_BY}. In this way
one can obtain globally valid data provided the initial black hole
separation, $r_{12}$, is within the PN regime. This is seen as a first step
away from the standard puncture approach and a step towards constructing
astrophysically realistic initial data. Using ADMTT gauge, the expressions
for the 3-metric and the extrinsic curvature are taken up to $O(v/c)^5$ and
$O(v/c)^4$ respectively, where $v\sim \sqrt{M/r_{12}}$ with $M=m_1+m_2$ and
$c$ is the speed of light. A formal expansion parameter, $\epsilon \sim
(v/c)$, is used below to distinguish terms of different PN order. The
3-metric in ADMTT gauge is given as
\begin{eqnarray}
\label{eq:g2PN_ADMTT}
g^{PN}_{ij} = \psi^4_{PN}\delta_{ij} + \epsilon^4 h^{TT}_{ij(4)}
+ O(\epsilon^5).
\end{eqnarray}
The full expression for $h^{TT}_{ij}$ has been computed by Kelly et
al.~\cite{Kelly:2007uc}. It denotes the gravitational wave contribution to
the 3-metric. The conformal factor 
\begin{eqnarray}
\psi_{PN} = 1 + \sum_{A=1}^2{
\epsilon^2\frac{m_A}{2r_A} 
+\epsilon^4 \frac{\frac{P^2_A}{2m_A}-\frac{m_1m_2}{r_{12}}}{2r_A}
+O(\epsilon^6)},
\end{eqnarray}
is very similar to that of standard punctures in Eqs.~\eqref{eq:psiBL} and
~\eqref{eq:psiPunc}. The only difference is that it contains an additional
term at order $O(v/c)^4$ and no $u$ piece as in Eq.~\eqref{eq:psiPunc}
since the constraints are satisfied only approximately. It must be noted that
the PN expressions in Eqs.~\eqref{eq:g2PN_ADMTT} and ~\eqref{eq:K2PN_ADMTT}
are not pure Taylor expansions in $(v/c)$ since terms such as $\psi^4$ in
Eq.~\eqref{eq:g2PN_ADMTT} are not expanded in powers of $\epsilon$. This
amounts to adding specific higher-order terms for the purpose of creating a
metric that is similar to the standard puncture expression. This selective
inclusion of terms does not improve the accuracy of the PN expansion,
however it is performed to ensure the presence of black hole apparent
horizons in the initial data~\cite{Tichy02}. The extrinsic curvature is
\begin{equation}
\label{eq:K2PN_ADMTT} 
 K^{ij}_{PN} = \psi^{-10}_{PN}\left[\epsilon^3 A^{ij}_{BY} -
\epsilon^5\frac{1}{2}\dot{h}^{TT}_{ij(4)}
-\epsilon^5{(\phi_{(2)}\tilde{\pi}^{ij}_{(3)})}^{TT}\right],
\end{equation}
where the leading order term contains the Bowen-York extrinsic curvature as
in the case of standard punctures. The additional terms at order
$O(v/c)^5$ can be found in~\cite{Tichy02}. The trace of the
extrinsic curvature can be shown to vanish up to $O{(v/c)}^6$.


\section{Asymptotic Matching of 4-Metrics: An Overview}
\label{Asymptotic Matching}
Another alternative to the standard puncture method involves a
post-Newtonian expansion of the 4-metric away from the black holes. This
expansion is then asymptotically matched to tidally perturbed black hole
metrics used for, and in close proximity to, the black holes.
This approach is believed to produce more
astrophysically realistic initial data based on several reasons; the method
is built on physical approximations and not on the potentially unphysical
ansatz of assuming the punctures form close to each black hole, and the level
of analytic control over the physical approximations allows for the method,
in principle, to accommodate ever higher-order expansions if desired.

\begin{figure}[htp]
\includegraphics[scale=0.30,clip=true]{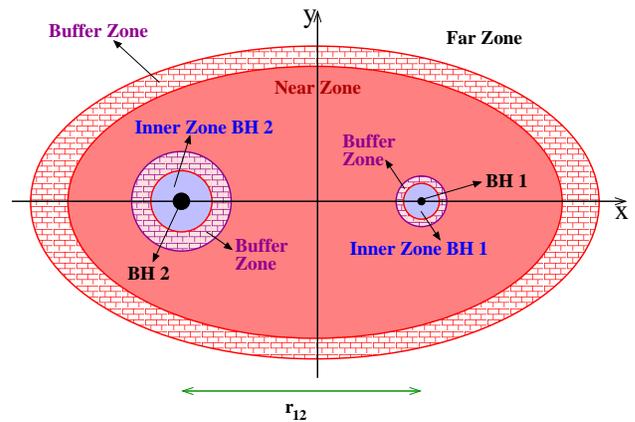}
\caption{\label{Zones}
This figure illustrates the separation of the spacetime into specific zones:
the inner zones, near zone and far zone. In each zone one can use a specific
approximation. The corresponding buffer zones are the regions where two zones
overlap.}
\end{figure}

Matching requires dividing the spacetime into 4 zones; 2 inner zones, a near
zone, and a far zone. The inner zones are considered as regions where 
perturbative Schwarzschild solutions are valid. The near zone is the region
where PN theory will hold (provided $({v}/{c})\ll 1$). The far zone is the
region of spacetime where retardation effects will influence the system.
Within each zone, 4-metric approximations are constructed in some coordinate
system with corresponding parameters (mass, momentum, etc.). Matching these
different approximations becomes possible because their regions of validity
overlap in certain regions called buffer zones. Figure~\ref{Zones} shows a
schematic sketch of the different regions. Asymptotic
matching~\cite{Burke:1970wx,D'Eath:1975qs,D'Eath:1975vw,Thorne:1984mz}
involves comparing two asymptotic solutions inside of a 4-volume. This
returns a map between the local coordinates and parameters of each of the
different regions, and forces both solutions to be asymptotic to each other
within the buffer zone. It is inside the buffer zone where these solutions
can be merged into a smooth global metric by way of a transition function.
We refer the reader 
to~\cite{Yunes:2005nn,Yunes:2006iw,JohnsonMcDaniel:2009dq} for our choice of
transition functions.

For a coordinate separation of $r_{12}$ the gravitational wavelength is
given by
\begin{eqnarray}
\lambda_{GW} = \pi\sqrt{\frac{r_{12}^3}{m}}.
\end{eqnarray}
A diagram of the regions is shown in Figure~\ref{Zones}, and a summary of the
regions' influences is seen in Table 1. 

\begin{table}[htp]
 \begin{center}
  \begin{tabular}{|| c | c | c || }
    \hline
    Zone & $r_{in}$ & $r_{out}$ \\ \hline
    Inner Zones & $0$ & $\ll r_{12}$ \\
    Near Zone & $\ll m_A$ & $\ll \frac{\lambda_{GW}}{2\pi}$ \\
    Far Zone & $\ll r_{12}$ & $\infty$ \\ \hline
    \hline
  \end{tabular}
 \end{center}
\caption{\label{Zone_Table}Location of inner and outer boundaries of each
spacetime zone.}
\end{table}

\subsection{Matching an ADMTT-Gauge Post-Newtonian Metric to a Black Hole
Perturbative Metric}
\label{ADMTT-match}
One of the first implementations that has used the above asymptotic
matching approach is described in~\cite{Yunes:2006iw}. In this initial data
scheme, a PN 4-metric in ADMTT
coordinates~\cite{Schaefer85,Jaranowski97,Jaranowski98a} is used in the near
zone. Since in this approach the Post-Newtonian metric is accurate only up
to terms of order $O(v/c)^2$, there are no retardation effects and the
near zone results are also valid in the far zone. In the inner zones, a
perturbed Schwarzschild metric in isotropic coordinates is used. A globally
valid 4-metric is then obtained by asymptotic matching. The matching
procedure is simplified by the fact that ADMTT and isotropic coordinates are
similar. The matched result has errors of $O(v/c)^4$ away from the
black holes and errors of order $O(v/c)^2 O(r_A/r_{12})^3$ near each black
hole. The PN part of the matched results that is valid away from the black
holes has the same 3-metric and extrinsic curvature as in
Eqs.~\eqref{eq:g2PN_ADMTT} and ~\eqref{eq:K2PN_ADMTT}, but with all terms of
order $\epsilon^4$ and higher dropped. The perturbed black hole solutions
that are valid near each hole can be found in~\cite{Yunes:2006iw}. Both
metrics are matched in the buffer zones and smoothed through the transition
function mentioned earlier.

\subsection{Matching a Harmonic-Gauge Post-Newtonian Metric to a 
Black Hole Perturbative Metric at Higher Order}
\label{Harmonic-match}
The second matching approach we consider here is from work by 
Johnson-McDaniel \textit{et al.}~\cite{JohnsonMcDaniel:2009dq}.
It applies higher-order asymptotic matching of the 4-metric solutions in
the buffer zones. For this approach the inner zones approximate the 4-metric
in a quasi-Cartesian form of Cook-Scheel harmonic 
coordinates~\cite{Cook97b}, the near zone
now approximates the 4-metric in harmonic coordinates, and the far zone
region is now considered through approximating the 4-metric in harmonic
coordinates. The reader is referred to~\cite{JohnsonMcDaniel:2009dq} for the
complete construction of the 4-metric within the near zone regime. The inner
zone solution is expressed to higher-order than what was used in the previous
matching approach in ADMTT gauge and is found in its full form in
~\cite{JohnsonMcDaniel:2009dq}. The inclusion of a far zone contribution
introduces an additional buffer zone, wherein another transition function
will smoothly join the near and far zone approximations in the region where
the errors from both approximations are comparable. 

The main difference between this matching approach and the one prior is that
it is now carried out to higher order; in both the PN expansion of the
4-metric away from the black holes and also the perturbation expansion
of the inner zone black hole solutions. Also, the PN expansions are now in
harmonic gauge. The matched result has errors of $O(v/c)^5$ away
from the black holes and errors of order $O(v/c)^5
O(r/r_{12})^3$ near each black hole. This result is then further
augmented by additional (but formally unmatched) PN terms away from the black
holes which include gravitational radiation, so that away from the black
holes one has only an error of $O(v/c)^6$. The choice of harmonic
coordinates is one of convenience; matching has already been established
between the near and far zones through~\cite{Blanchet:2006zz} which 
simplifies the inclusion of the far zone region. This high-order
matching yields more accurate results than the previous lower-order matching
in ADMTT gauge.
With the inclusion of initial gravitational radiation, we can
expect improvements in the data set over the standard punctures approach.

In~\cite{Chu2012a} the tidal deformations present in the inner zones
of~\cite{JohnsonMcDaniel:2009dq} (but not the near and far zones) 
are added to a superposition of two spinning black holes in Kerr-Schild
coordinates. Evolutions display a reduction in the (2,0) mode of the outgoing
gravitational waves. The expectation is that a full inclusion of outer zone
phenomena, as in our work, will also reduce the (2,2) mode spurious
signal.


\section{Numerical Evolution Results}
\label{Numerical_Explanantion_&_Results}
Evolutions from all initial data were done using the
{\fontfamily{pcr}\selectfont BAM}
code~\cite{Bruegmann:2003aw,Husa:2007hp,Marronetti:2007wz,Brugmann:2008zz}.
The gravitational fields are evolved using the
BSSNOK formalism~\cite{Nakamura87,Shibata95,Baumgarte:1998te} in the
variation known as the ``moving punctures" method~\cite{Campanelli:2005dd,
Baker:2005vv}. Thus the $3$-metric $g_{ij}$ is written as
\begin{equation}
g_{ij} = \chi^{-1} \tilde{\gamma}_{ij}
\end{equation}
where the conformal metric $\tilde{\gamma}_{ij}$ has unit determinant.
In addition, the extra variable
\begin{equation}
\label{eq:Gammatilde}
\tilde{\Gamma}^i
= \tilde{\gamma}^{ij} \tilde{\gamma}^{kl} \tilde{\gamma}_{jk,l}
\end{equation}
is introduced where $\tilde{\gamma}^{ij}$ is the inverse of
the conformal metric.
Furthermore, the extrinsic curvature is split into its trace-free part,
$\tilde{A}_{ij}$, and its trace, $K$, which is given by  
\begin{equation}
K_{ij} = \chi^{-1}
         \left( \tilde{A}_{ij} + \frac{K}{3} \tilde{\gamma}_{ij} \right) .
\end{equation}
The particulars of our BSSNOK
implementation can be found in~\cite{Brugmann:2008zz}.

All our evolutions use equal-mass non-spinning black hole binaries at
initial separation of $10M$. Also, all sets are tested through evolutions
using a resolution of $\sim{3M}/{224}$ near the black holes. Furthermore,
all sets use the same expressions for $\alpha$ and $\beta^i$ at the initial
time;
\begin{eqnarray}
\alpha &=& {\psi_{BL}}^{-2},\\
\beta^i &=& 0 .
\end{eqnarray}
To evolve the lapse and shift we use the ''1+log'' slicing
condition~\cite{Bona94b} modified with the addition of an advection term as
used in~\cite{Baker:2005vv,vanMeter:2006vi}, and the ``gamma freezing
condition''~\cite{Alcubierre02a} modified and used
in~\cite{Brugmann:2008zz,vanMeter:2006vi}, where it is labeled as the
``shifting-shift case'' in the former and the ``000'' shift choice in the
latter. For the initial data obtained by matching or from PN approaches, the
constraints are only approximately satisfied. In the case of standard punctures
the constraints are solved numerically using the approach
in~\cite{Ansorg:2004ds}.

\subsection{Filling Black Hole Interiors}
\label{BHfiller}
In the case of initial data obtained by asymptotic matching, each black
hole is described by a perturbed black hole solution. This black hole solution
can contain physical singularities inside the event horizons that have to be
dealt with before a numerical evolution can be attempted. We have developed
an algorithm (called {\fontfamily{pcr}\selectfont BHfiller} in our code)
that fills a specified region inside the black hole apparent horizon with
smooth data. The justification for this procedure is that the physics
outside a black hole remains unaffected if we only change the inside of the
horizon. Our algorithm modifies the BSSNOK variables at the initial time as
follows: For each black hole (centered at the point $\vec{x}_{A}$) 
we pick a sphere of radius $r_{fill}$ contained inside the black
hole horizon. The radius $r_{fill}$  is chosen such that this sphere
contains all singular points. To set valid data at each point $\vec{x}$
inside this
sphere we use a weighted average of linear extrapolation and a
value that corresponds to standard puncture data. 
Let us define $r=|\vec{x}-\vec{x}_{A}|$, 
$\hat{n} = (\vec{x}-\vec{x}_{A})/r$ and denote a particular BSSNOK variable
component at point $\vec{x}$ by $u(\vec{x})$.
If $r<r_{fill}$ we set 
\begin{equation}
u(\vec{x}) = u_r \varUpsilon + u_0 (1 - \varUpsilon),
\end{equation}
where 
\begin{equation}
u_r = u(r_{fill}\hat{n}) + [\partial_r u(r_{fill}\hat{n})](r-r_{fill}),
\end{equation}
\begin{eqnarray}
u_0 = \begin{cases}
       0.1&	\text{if $u$ is $\alpha$},\\
      \left(2+\dfrac{m_{A}}{2r}\right)^{-4}&	\text{if $u$ is $\chi$},\\
       1&	\text{if $u$ is $\tilde{\gamma}_{xx}$, $\tilde{\gamma}_{yy}$
or $\tilde{\gamma}_{zz}$},\\
       0&	\text{otherwise}.
      \end{cases}
\end{eqnarray}
and the weight is given by 
\begin{eqnarray}
\varUpsilon &=&
\frac{1}{2}\left\{1+\tanh\left[\frac{48}{125}
\left(\frac{r_{fill}}{r_{fill}-r}-\frac{3~r_{fill}}{2~r}\right)
\right]\right\}.
\end{eqnarray}

We apply this method of filling the region inside radius $r_{fill}$ to all
BSSNOK variables except $\tilde{\Gamma}^i$, which we simply recompute
from the filled $\tilde{\gamma}_{ij}$ using Eq.~\eqref{eq:Gammatilde}.
We only need to use this algorithm for the case of asymptotically matched
data sets, as all other data sets are free of any physical singularities.

Note that the constraints are generically not satisfied in the filled region.
We have tested our filling algorithm by evolving standard puncture data with
filling, applied at the initial time or some predetermined later time, and
without filling and compared various quantities. Gauge invariant quantities,
such as the gravitational wave amplitude as a function of gravitational wave
frequency, are unchanged when we compare with and without black hole filling.
We have also checked that in our evolutions no visible constraint violations
are emitted by the black holes filled with this algorithm. The latter is
expected since the BSSNOK system together with the gauge conditions used here
has been shown~\cite{Brown:2007pg} to lead to causal constraint propagation,
so that any constraint violations introduced inside the black holes cannot
affect the exterior spacetime. Notice, however, that the BSSNOK system has
superluminal gauge modes~\cite{Brown:2007pg}. Thus gauge dependent quantities
such as the lapse are somewhat different, even outside the black holes. Hence
the black hole trajectories (while qualitatively the same) are slightly
different as well.

\subsection{Shortcomings of Low-Order Asymptotic Matching}
\label{Bad_ADMTT-match}
\begin{figure}[htp]
\includegraphics[scale=0.33,clip=true,angle=-90]
{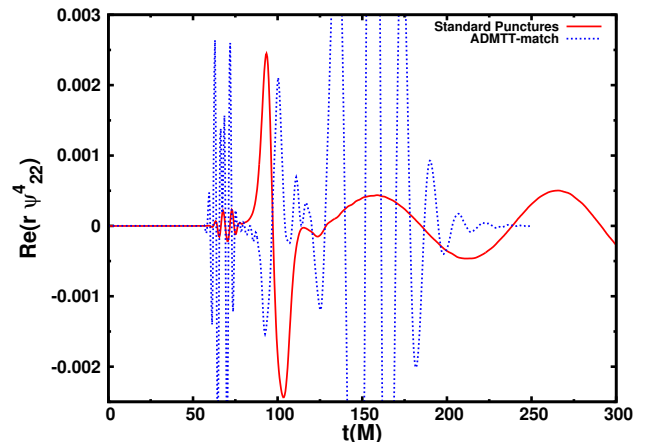}
\caption{\label{Compare_cyPunc_v_ADMMTT-match_Re_rpsi4}
The (2,2) mode of $\psi_4$ extracted at a radius of $r=90M$ for standard
punctures and ADMTT-match initial data sets. Both waveforms show no initial
wave signature at $t=0$, due to their reliance on conformal flatness for the
3-metric expressions. It can also be seen that ADMTT-match merges much faster
(within $250M$). This is evidence that the orbits were not quasicircular at
all.
}
\end{figure}

The real part of the dominant $\ell=2$, $m=2$ waveform mode of $\psi_4$
illustrates whether a gravitational wave signal has been implemented starting
from the $t=0$ slice. Figure~\ref{Compare_cyPunc_v_ADMMTT-match_Re_rpsi4}
shows the waveforms, extracted at radius $r=90M$, for a simulation starting
from standard puncture data and that produced from the first data set;
asymptotic matching of a low-order PN expansion in ADMTT-Gauge to a
perturbative black hole solution (hereafter called ADMTT-match).
As with standard punctures, ADMTT-match fails to incorporate gravitational
radiation at $t=0$. This is not surprising as
the 3-metric is still conformally flat. 

\begin{figure}[htp]
\includegraphics[scale=0.33,clip=true,angle=-90]{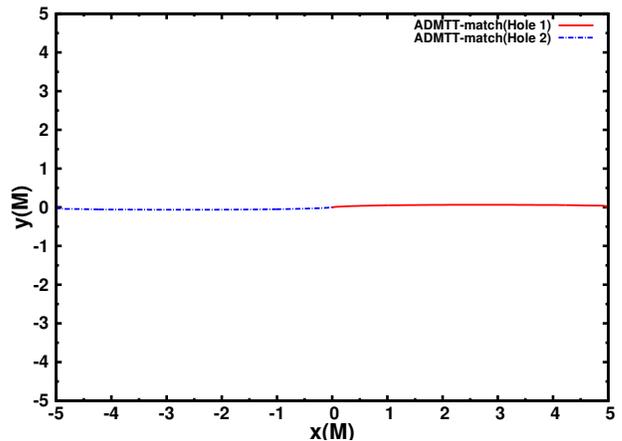}
\caption{\label{ADMTT-match_Trajectories}
Trajectories of both black holes in the ADMTT-match approach. A nearly
head-on collision between the two holes indicates there is insufficient
tangential momenta to satisfy a PN quasicircular orbit starting from $t=0$.}
\end{figure}

Another undesired result is that the binary merges within a time of $250M$,
which is much too fast. This behavior indicates an accelerated evolution
timescale that is not expected for a quasicircular inspiral. Figure
~\ref{ADMTT-match_Trajectories} shows the orbit of the binary system in the
ADMTT-match approach.
We see that ADMTT-match leads to nearly head-on trajectories for the black
holes. The individual black holes do not have enough momentum to complete
even one orbit. One can track this problem back to the order of asymptotic
matching used in the construction of these initial data. The terms in the
extrinsic curvature that are associated with momentum are computed from the
4-metric perturbation expansions used near the black holes. However, as
discussed in the Appendix, the matching procedure used in~\cite{Yunes:2006iw} is
of an order that is too low to correctly obtain the the $O(v/c)^3$ piece of the
extrinsic curvature that would lead to an orbiting black hole. Thus we see that
this data set is merely a proof of concept. To obtain useful initial data, we
need to use higher order matching.

\subsection{Analysis of Higher-Order Methods}
\label{Harmonic-match_&_ADMTT-PN}

In this subsection, we compare evolutions starting from data obtained by the
higher order matching approach discussed in Sec.~\ref{Harmonic-match} and
data from the pure PN approach of Sec.~\ref{ADMTT-PN} with standard puncture
initial data. However, we first need to inform the reader of a modification
to the transition function in~\cite{JohnsonMcDaniel:2009dq} used for
smoothing the far-zone and near-zone metric contributions in the outermost
buffer zone. The parameters $r_0$ and $w$, which define where the transition
begins and the width of the transition window respectively, have been
adjusted. The analysis in~\cite{JohnsonMcDaniel:2009dq} 
uses $r_0=\lambda_{GW}/5$ and $w=\lambda_{GW}$ [see Eqs.~(8.4) and (8.6)
in~\cite{JohnsonMcDaniel:2009dq}],
which then put the midpoint (where the transition function has a value of
1/2) at $r\approx14\lambdabar_{GW}$. In the work here we 
choose values of $r_0=0.044 \lambda_{GW}$ and $w=0.22 \lambda_{GW}$,
which lead to a midpoint at
$r\approx\lambdabar_{GW}$. This adjustment is justified because one would
expect the transition region (from near zone to far-zone) to be more
suitable closer to $\lambdabar_{GW}$ than $14\lambdabar_{GW}$.


\subsubsection{Waveforms}
\label{Waveforms}
We first discuss the gravitational waveforms obtained from these data sets.
In the figures that follow we denote our results as Harmonic-match or
ADMTT-PN; depending on whether we evolve initial data constructed by
asymptotic matching of a second-order post-Newtonian 4-metric expansion in
Harmonic gauge to a perturbative black hole solution, or data from a 
post-Newtonian expansion in ADMTT gauge. We do not expect to see the same
deficiencies as seen in ADMTT-match.
\begin{figure}[htp]
\includegraphics[scale=0.33,clip=true,angle=-90]
{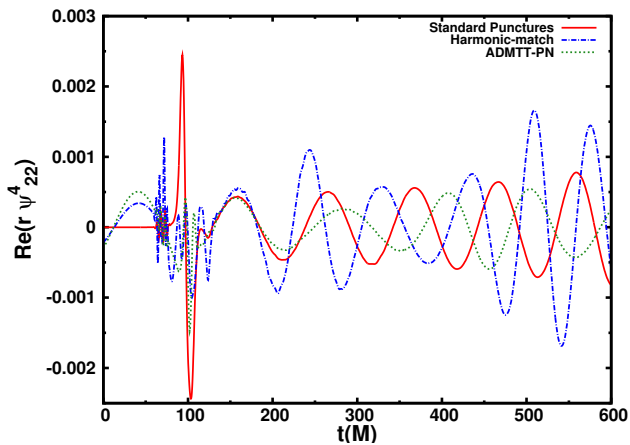}
\caption{\label{Compare_cyPunc_v_Harmonic-match_v_ADMTT-PN_Re_rpsi4}
The (2,2) mode of $\psi_4$ extracted at a radius of $r=90M$ for standard
punctures, Harmonic-match, and ADMTT-PN data sets.
}
\end{figure}
Figure~\ref{Compare_cyPunc_v_Harmonic-match_v_ADMTT-PN_Re_rpsi4} 
shows the real part of the ($\ell=2$,$m=2$) waveform mode of $\psi_4$ at an
extraction radius of $90M$ for all three data sets. Harmonic-match produces
gravitational waves starting from $t=0$, more characteristic of an
astrophysically realistic evolution. The diminished amplitude in the
spurious radiation region for Harmonic-match is an additional positive
outcome for this data set (although the ultimate goal is to completely
eliminate this behavior from the waveform). 
Figure~\ref{Compare_cyPunc_v_Harmonic-match_v_ADMTT-PN_Re_rpsi4} 
also shows the waveform for ADMTT-PN and standard punctures. As with
Harmonic-match, ADMTT-PN shows a non-zero waveform signature at $t=0$
and comparable reduction of the spurious radiation amplitude. It is
also noted that a comparative` reduction of the spurious radiation amplitude
has been seen in~\cite{Mundim:2010hu} as well.

\subsubsection{Eccentricity and Horizon Mass}
\label{Eccentricity_&_Mass}
\begin{figure}[htp]
\includegraphics[scale=0.33,clip=true,angle=-90]
{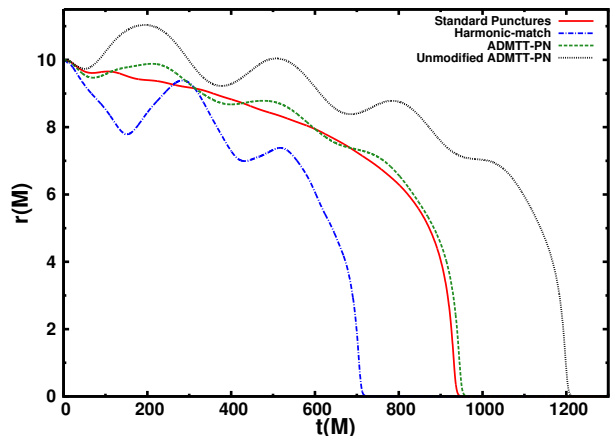}
\caption{\label{Compare_cyPunc_v_Harmonic-match_v_ADMTT-PN_Separation}
Coordinate separation for standard punctures, Harmonic-match, unmodified
ADMTT-PN, and modified ADMTT-PN.}
\end{figure}

We can see from
Fig.~\ref{Compare_cyPunc_v_Harmonic-match_v_ADMTT-PN_Re_rpsi4} that the
waveform behavior of both Harmonic-match and ADMTT-PN is non-monotonic in 
amplitude. This differs from the standard chirp signal expected during
the inspiral phase of the two black holes. The reason for this
behavior is due to the black hole orbits being noticeably eccentric. The high
eccentricity can be seen through plots of the coordinate separation over time
(See Fig.~\ref{Compare_cyPunc_v_Harmonic-match_v_ADMTT-PN_Separation}).
For a sense of scale, the coordinate separation for standard puncture
initial data is included to indicate the small level of deviation for an
evolution to be accepted as having low eccentricity.
Note, however, that unlike
in the case of standard punctures, the tangential momentum of the
black holes has not been fine-tuned and is simply coming from PN expressions
for circular orbits.
For example, in the case of ADMTT-PN we use the momentum for circular PN
orbits given by
\begin{eqnarray}
\label{eq:p_2PN_circ}
{p_{PN}}^2 &=&
\mu\left[\left[{M\Omega}\right]^{\frac{1}{3}}+\epsilon^2 \frac{
(15-\eta)M\Omega}{6} \right. \nonumber \\
& &\left.+\epsilon^4\frac{(441-324\eta-\eta^2)
{[M\Omega]}^{\frac{5}{3}}}{72} \right],
\end{eqnarray}
where $M=m_1+m_2$, $\mu = m_1 m_2/M$, $\eta=\mu/M$ and $\Omega$ is defined as
\begin{equation}
\label{eq:OmegafromD}
\Omega = 
\frac{1}{M}\left[\dfrac{64(\frac{r}{M})^3}{(1+2(\frac{r}{M}))^6} +
\dfrac{\eta}{(\frac{r}{M})^4} \right.
\left.+
\dfrac{\left(-0.625\mu+\eta^2\right)}{(\frac{r}{M})^5}\right]^\frac{1} {2}
\end{equation}

\begin{figure}[htp]
\includegraphics[scale=0.33,clip=true,angle=-90]
{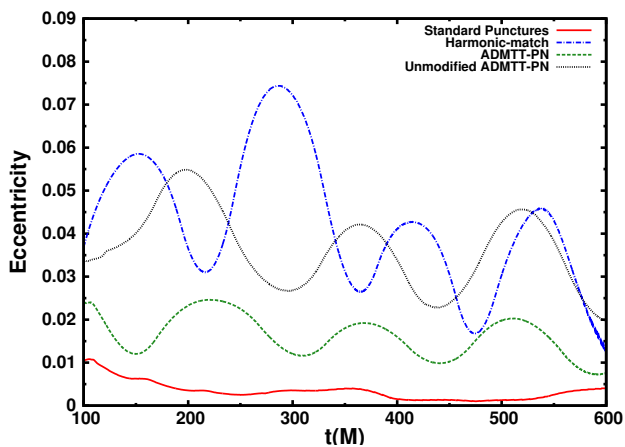}
\caption{\label{Compare_cyPunc_v_Harmonic-match_v_ADMTT-PN_Eccentricity}
Eccentricity estimation for standard punctures, Harmonic-match, unmodified
ADMTT-PN, and modified ADMTT-PN. The range between $100M$ and $600M$ is the
range to analyze all sets simultaneously while guaranteeing no
errors due to incomplete information (Before $100M$ our eccentricity estimator
which averages over an orbital time is unreliable,
and after $600M$ most sets have begun
merger so that one cannot define an eccentricity.).}
\end{figure}

Increased eccentricity, with respect to standard punctures, is again apparent
in Fig.~\ref{Compare_cyPunc_v_Harmonic-match_v_ADMTT-PN_Eccentricity}
for both the Harmonic-match and ADMTT-PN sets.
The eccentricity calculation is performed as in~\cite{Tichy:2010qa}, and is
calculated through a time-average over the orbital period. 

For the case of ADMTT-PN, we have added an additional parameter, $C_3$, that
can be used to adjust the initial tangential momenta of the black holes. The
modified tangential momentum, $p$ is given by
\begin{eqnarray}
\label{eq:ptang-corr}
p^2 = {p_{PN}}^2 + \epsilon^6
\frac{{\mu}^2 M}{8r_{12}}C_3\left(\frac{M}{r_{12}} \right)^3 ,
\end{eqnarray}
Note that the term with $C_3$ comes into the expression at order
$O(v/c)^6$. Thus this additional term is of the order of one of the PN
terms that were neglected in the expansion in Eq.~\eqref{eq:p_2PN_circ}. 
By adjusting $C_3$ we can then decrease the eccentricity. For
$C_3=-396.4844$ we change our tangential momentum to 98.3\% of the
original $p_{PN}$. This choice effectively minimizes the eccentricity.
In Figs.~\ref{Compare_cyPunc_v_Harmonic-match_v_ADMTT-PN_Separation}
and~\ref{Compare_cyPunc_v_Harmonic-match_v_ADMTT-PN_Eccentricity} 
we can compare the results with and without this modification
to Harmonic-match and standard punctures.

\begin{figure}[htp]
\includegraphics[scale=0.33,clip=true,angle=-90]
{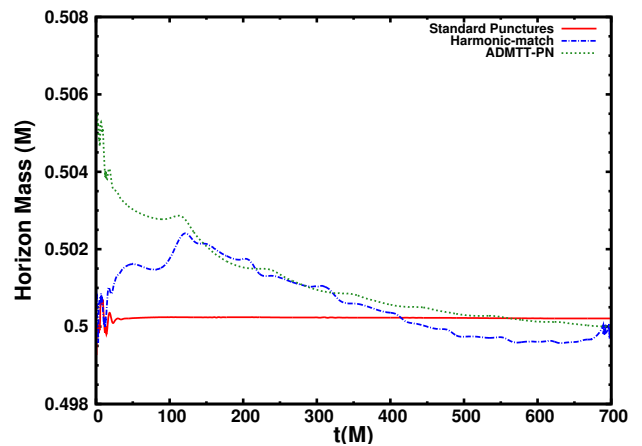}
\caption{\label{Compare_cyPunc_v_Harmonic-match_v_ADMTT-PN_Horizon0}
Apparent horizon mass of one of the black holes in the inspiral phase 
under the standard punctures, Harmonic-match, and
ADMTT-PN approach.}
\end{figure}

A useful local definition for the mass of a black hole is the apparent
horizon mass. An algorithm for finding apparent horizons without any
symmetries is described in~\cite{Gundlach97a}. Over time, numerous variations
on Gundlach's methodology have produced routines for many of the main
computational groups involved in black hole simulation. The version used here
is known as {\fontfamily{pcr}\selectfont AHmod} and was developed by Norbert
Lages~\cite{Lages2010a} as an improvement to the CACTUS thorn
{\fontfamily{pcr}\selectfont AHfinder} developed by Miguel
Alcubierre~\cite{Alcubierre98b}.
Figure~\ref{Compare_cyPunc_v_Harmonic-match_v_ADMTT-PN_Horizon0} shows the
apparent horizon mass of one of the (equal mass) black holes for all three
data sets versus time. We see that ADMTT-PN and Harmonic-match show much more
variation than standard puncture data. We see that the two alternative
methods lead to a ``mass loss'' over the evolution time where they eventually
approach the apparent horizon mass  obtained from evolving standard puncture
data.
This ``mass loss'', albeit of low scale, 
is unphysical and we expect it is due to larger Hamiltonian constraint
violations in the approximate initial data sets. For example,
in~\cite{Mundim:2010hu} the Hamiltonian constraint (but not the momentum
constraint) is solved numerically for ADMTT-PN, which leads to a horizon
mass that is closer to the result for standard puncture data.

\begin{figure}[htp]
\includegraphics[scale=0.33,clip=true,angle=-90]
{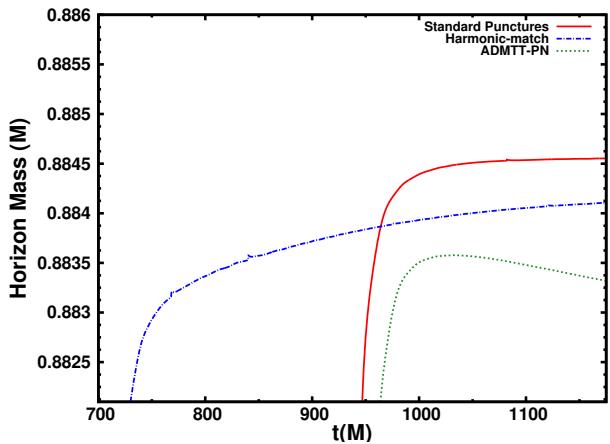}
\caption{\label{Compare_cyPunc_v_Harmonic-match_v_ADMTT-PN_Horizon2}
Apparent horizon masses of the final black hole after merger 
under the standard punctures, Harmonic-match, and
ADMTT-PN approach.}
\end{figure}

Figure~\ref{Compare_cyPunc_v_Harmonic-match_v_ADMTT-PN_Horizon2} shows the 
apparent horizons masses of the final black hole after merger for evolutions
with each data set. The masses obtained with the alternatives tend to deviate
from punctures, with Harmonic-match yielding an increase and ADMTT-PN
yielding a decrease in the common horizon mass respectively over time.
This decrease is certainly unphysical and is likely
related to larger constraint violations.

\subsubsection{Constraints}
\label{Constraints}

\begin{figure}[htp]
\includegraphics[scale=0.33,clip=true,angle=-90]
{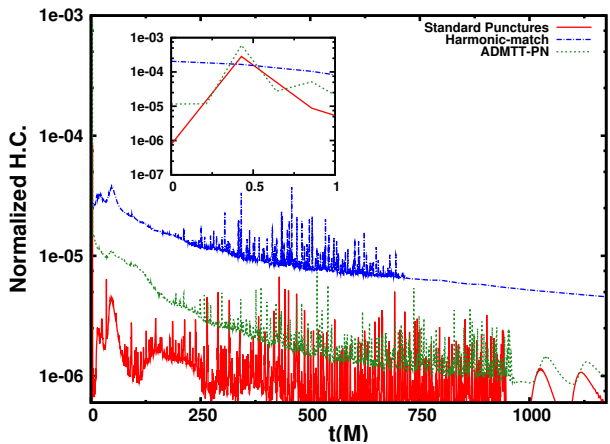}
\caption{\label
{Compare_cyPunc_v_Harmonic-match_v_ADMTT-PN_HC}
The $L^2$-norm of the Hamiltonian constraint violation for the standard
punctures, Harmonic-match, and ADMTT-PN approach. The y-axis is plotted in
$log_{10}$. The inset shows the violations at the beginning of the
evolution.
}
\end{figure}

The violation of the Hamiltonian constraint equation at the beginning of the
evolution, as seen in
Fig.~\ref{Compare_cyPunc_v_Harmonic-match_v_ADMTT-PN_HC} is between 2 and 3
orders of magnitude larger for Harmonic-match and
ADMTT-PN than for standard punctures at $t=0$. We expect that these violations
are responsible for the observed drift in the apparent horizon masses.
Over time, the evolution of ADMTT-PN relaxes the constraints
to the regime of standard punctures where both are almost indistinguishable.
This result brings the expectation that these initial data methods are stable
over the life of the evolution, and a case for astrophysical relevance.

\begin{figure}[htp]
\includegraphics[scale=0.33,clip=true,angle=-90]
{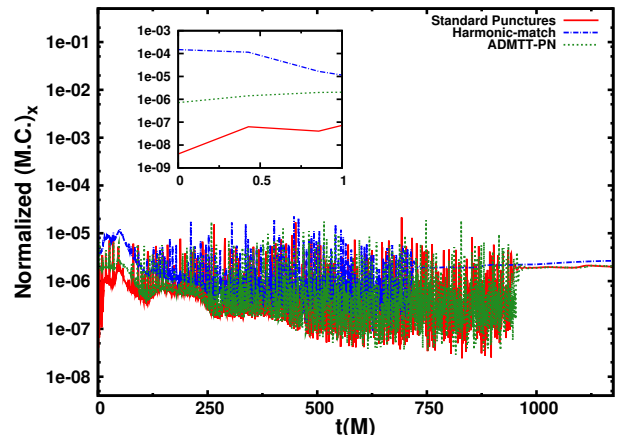}
\caption{\label
{Compare_cyPunc_v_Harmonic-match_v_ADMTT-PN_MC}
The $L^2$-norm of the $x$-component of the Momentum constraint violation,
for the standard punctures, Harmonic-match, and ADMTT-PN approach.
The y-axis is plotted in $log_{10}$.
The inset shows the violations at the beginning of the evolution.
}
\end{figure}

Figure~\ref{Compare_cyPunc_v_Harmonic-match_v_ADMTT-PN_MC}
shows the $L^2$-norm of the $x$-component of the Momentum constraint.
It is seen that the initial discrepancies (compared to standard punctures)
are of 2 orders of magnitude for ADMTT-PN and roughly 5 orders of magnitude for
Harmonic-match.
The $y$ and $z$ components are so similar that they would yield 
almost the same plots. 
Over time, all data sets relax to within an order of
magnitude difference until they become almost indistinguishable during
the merger and ringdown phases.


\section{Discussion}
\label{Discussion}

We have investigated three alternative approaches of constructing initial
data for a binary black hole system. Two of these alternatives involve
asymptotic matching of post-Newtonian expansions to perturbative black hole
solutions of the metric, and one uses a resummed post-Newtonian expansion of
the metric everywhere. We numerically evolve all three alternatives using
the BSSNOK system~\cite{Baumgarte:1998te} together
with the moving punctures approach~\cite{Campanelli:2005dd,Baker:2005vv}
and compare them with evolutions that start from standard
puncture initial data. The two alternative data sets that were derived by
asymptotic matching have not been evolved before. Since they contain
singularities near the black hole centers, we have also developed a
particular algorithm to fill a certain region inside each black hole with
smooth data that approaches puncture data at the center. In this way we can
avoid the use of black hole excision, so that we are able to use the same
moving punctures approach to evolve all our data sets.

We find that the simplest data set, using low-order asymptotic matching, does
not lead to physically accurate evolutions. Insufficient expansion of the
extrinsic curvature leads to black hole momenta that are far too small for
quasicircular orbits, so that we observe an almost head-on collision.
Also, low-order expansion of the metric yields gravitational waveforms
lacking an initial signature. Thus, this initial data set is not an
improvement over standard punctures.

Higher-order expansion techniques resulted in the production of gravitational
wave signals at the initial time as well as proper quasicircular inspiral
orbits that subsequently lead to merger and ringdown. Eccentricity measures
were higher for the Harmonic-match and ADMTT-PN sets than what is seen in
standard punctures, when the puncture momentum parameters are fine-tuned to
achieve low eccentricities. However, by tuning our 3PN order $C_3$ parameter
(see Eq.~\eqref{eq:ptang-corr}) we were able to bring the eccentricity of
ADMTT-PN down as well. Both Harmonic-match and ADMTT-PN sets look promising
in the sense that, unlike for standard puncture data, they have realistic
gravitational wave signals built-in from the beginning of the evolution.
In addition, they show a reduction of spurious radiation when we compare
with standard puncture data.

There are, however, still problems with these alternative data sets
that stem from the fact that they only approximately satisfy the constraint
equations of general relativity. We have monitored the Hamiltonian and
momentum constraint violations during the evolutions. We find that
constraint violations of Harmonic-match and ADMTT-PN sets are above the
usual level coming from numerical truncation errors observed when we evolve
standard puncture data. In the case of ADMTT-PN these violations eventually
decrease and reach a level similar to what is seen for standard puncture
data evolutions. However, until this happens we see an unphysical drop in the
apparent horizon masses. In the case of Harmonic-match, the constraint
violations start at an even higher level and never quite fall to the level
observed for standard puncture data evolutions, and again result in
unphysical behavior of the apparent horizon masses.

Future improvements need to be focused on these constraint violations.
An obvious remedy would be to project the alternative data sets onto
the solution manifold of general relativity by solving the constraint
equations, e.g. as in~\cite{Tichy02} by using the York-Lichnerowicz
conformal decomposition~\cite{York73}. If the constraints are satisfied
we expect the apparent horizon masses to be strictly non-decreasing over
time.
Evolutions, horizon masses, and eccentricity could then be compared to the
standard punctures approach and conclusions could be made on more stringent
grounds.

Most astrophysical black holes are expected to be spinning. It is thus
important to include spin in the initial data construction. As already
mentioned in Sec.~\ref{Standard Punctures}, standard punctures contain spin
parameters that enter the Bowen-York extrinsic curvature in
Eq.~(\ref{eq:A_BY}), so that one can set up initial data with spins.
Including spins in the alternative data sets discussed in this paper is
possible as well. For any of the data sets that are produced from asymptotic
matching one would have to redo the matching calculation as
in~\cite{JohnsonMcDaniel:2009dq} and match a post-Newtonian 4-metric with
spin to a perturbed black hole metric with spin. This calculation, while
certainly long and tedious, is in principle possible. The situation is much
simpler for the post-Newtonian data set in ADMTT gauge discussed in
Sec.~\ref{ADMTT-PN}. To leading order, the spin only enters the Bowen-York
piece $A_{BY}^{ij}$ in Eq.~(\ref{eq:K2PN_ADMTT}) in the same way as in the
case of standard punctures. It would thus be easily possible to add these
missing spin terms. This alone would already result in spinning black holes.
However, in order to obtain the correct waveform at the initial time we have
to ensure that $h^{TT}_{ij(4)}$ in Eqs.~(\ref{eq:g2PN_ADMTT}) and
(\ref{eq:K2PN_ADMTT}) is computed for post-Newtonian particle trajectories
where spin is included in the equation of motion. Since post-Newtonian
trajectories for spinning particles are well known, and since
$h^{TT}_{ij(4)}$ is known for arbitrary trajectories~\cite{Kelly:2007uc}, the
inclusion of spin in this data set is not a complicated problem.


\begin{acknowledgments}
This material is based upon work supported by the National Science Foundation
under Grants PHY-0855315, PHY-1204334 and DGE: 0638662. 
Computational resources were
provided by the Kraken cluster (allocation TG-PHY090095) at the National
Institute for Computational Sciences. Any opinions, findings, and conclusions
or recommendations expressed in this material are those of the authors and
do not necessarily reflect the views of the National Science Foundation.
\end{acknowledgments}

\appendix

\section{Drawbacks of Asymptotic Matching}
\label{LowMatchProblem}

In~\cite{Yunes:2005nn,Yunes:2006iw,JohnsonMcDaniel:2009dq} the ansatz,
\begin{equation}
X^\alpha(x^\beta) = 
 \sum_{j=0}^{5}\left(\frac{m_2}{b}\right)^{j/2}\od{X^\alpha}{j}(x^\beta)
 + O(v/c)^6
\end{equation}
is made for the coordinate transformation that transforms the 
coordinates $X^\alpha$ of the perturbed black hole solution in
inner zone 1 to the coordinates $x^\beta$
that are used in the near zone for the post-Newtonian metric.
Note that $\left(\frac{m_2}{b}\right)^{j/2}\sim O(v/c)^j$.
In Sec.~V of~\cite{JohnsonMcDaniel:2009dq} it is shown that 
the zeroth order piece $\od{X^\alpha}{0}$ of this coordinate transformation
corresponds to a simple translation and the first order piece is
\begin{equation}
\od{X_\alpha}{1} = \od{F_{\beta\alpha}}{1} x^\beta + \od{C_\alpha}{1},   
\end{equation}
where $\od{F_{\beta\alpha}}{1}$ is anti-symmetric. The
$\od{F_{\beta\alpha}}{1}$ and $\od{C_\alpha}{1}$ correspond to constants of
integration that appear when matching at $O(v/c)$.
As shown in Sec.~V.E. of~\cite{JohnsonMcDaniel:2009dq}, the
$\od{F_{\beta\alpha}}{1}$ can only be determined by matching 
the two 4-metrics up to $O(v/c)^3$. To be precise,
$\od{F_{\beta\alpha}}{1}$ comes from matching the 
$g_{ij}$ and $g_{00}$ components at $O(v/c)^3$. 
This pattern continues in the sense that
at each order of matching one finds free constants of integration that can
only be fixed once matching has been performed two orders higher in $v/c$.

In the case of the lower order matching procedure used 
in~\cite{Yunes:2005nn,Yunes:2006iw}, matching of the $g_{ij}$ and $g_{00}$
components of the two 4-metrics has only been performed up to 
$O(v/c)^2$. This means that the piece $\od{F_{\beta\alpha}}{1}$
could not be computed. In fact, it was merely set to a value
that simplified further calculations. This has important consequences.
As shown in Sec.~V.E. of~\cite{JohnsonMcDaniel:2009dq} the only non-zero
components of $\od{F_{\beta\alpha}}{1}$ are 
$\od{F_{02}}{1} = -\od{F_{20}}{1} = -\sqrt{m_2/m}$. This yields a term
$\left(\frac{m_2}{b}\right)^{1/2} \od{F_{02}}{1} t 
= -\frac{m_2}{m} \left(\frac{m}{b}\right)^{1/2} t$ in $Y=X^2(x^\beta)$,
which corresponds to a boost. This boost is needed to transform the metric
of the perturbed black hole solution in inner zone 1 (from a frame where it
is at rest to a frame where it is moving in the $y$-direction).
In the case of lower order matching as 
in~\cite{Yunes:2005nn,Yunes:2006iw} this term is absent, so that the metric 
used near the black hole is missing this boost. The resulting extrinsic
curvature (which is always smaller than the 3-metric by a factor of $v/c$) is
thus correct only up to $O(v/c)^2$ and has errors of $O(v/c)^3$, where the
missing boost would enter. This leads to a black hole without sufficient
momentum.

In general, this implies that if we want to obtain a 4-metric that is correct
up to $O(v/c)^n$ with the matching procedure
in~\cite{Yunes:2005nn,Yunes:2006iw,JohnsonMcDaniel:2009dq} we need to find
the matching coordinate transformation up to order $O(v/c)^{n+2}$.

\bibliography{references}

\end{document}